\documentclass[aps,prd,twocolumn,reprint,nofootinbib,superscriptaddress]{revtex4}

	\usepackage[utf8]{inputenc}
	\usepackage{amssymb}
	\usepackage{amsmath}
	\usepackage{graphicx}
	\usepackage{subfigure}
	\usepackage{dsfont}
	\usepackage{mathrsfs}
	\usepackage{natbib}       
	\usepackage[dvips]{color} 
\newcommand{\fix}[1]			
             {{\color{magenta} #1 }}

\begin{document}

\title{Gravothermal Catastrophe with a Cosmological Constant}
\author{Minos Axenides}
\affiliation{Institute of Nuclear and Particle Physics, National Center for Scientific Research Demokritos, GR-15310 Athens, Greece}
\author{George Georgiou}
\affiliation{Institute of Nuclear and Particle Physics, National Center for Scientific Research Demokritos, GR-15310 Athens, Greece}
\author{Zacharias Roupas}
\email[Corresponding author: ]{roupas@inp.demokritos.gr}
\affiliation{Institute of Nuclear and Particle Physics, National Center for Scientific Research Demokritos, GR-15310 Athens, Greece}
\affiliation{Physics Department, National Technical University of Athens, GR-15780, Athens, Greece}

\begin{abstract}
	We investigate the effect of a cosmological constant on the gravothermal catastrophe in the Newtonian limit. A negative cosmological constant acts as a thermodynamic `destabilizer'. The Antonov radius gets smaller and the instability occurs, not only for negative but also for positive energy values. A positive cosmological constant acts as a `stabilizer' of the system, which, in this case, exhibits a novel `reentrant behaviour'. In addition to the Antonov radius we find a second critical radius, where an `inverse Antonov transition' occurs; a series of local entropy maxima is restored.
\end{abstract}

\maketitle

\section{Introduction}
Antonov's gravothermal instability is an important effect in gravitational thermodynamics \cite{Antonov,Bell-Wood,Padman,Chavanis}. It has served as the prime paradigm for an extensive research into the
statistical mechanics of systems with long range interactions in different fields of physics \cite{Bell, Dauxois}, while phase transitions of self-gravitating systems is an active research field \cite{Chavanis2}. One may ask if and how this phenomenon depends on the vacuum background. A positive cosmological constant is nowadays one of the main candidates for dark energy \cite{Padman2,Harvey}, while negative cosmological constant has attracted a lot of attention due to the anti-de Sitter/Conformal Field Theory correspondence (AdS/CFT \cite{AdSCFT}). In addition, the stability properties of anti-de Sitter spacetime have become recently a subject of topical interest \cite{AdS1,AdS2}. We present here how gravothermal catastrophe is affected by a cosmological constant, i.e. by a de Sitter or anti-de Sitter vacuum. 
For convenience we call the non-relativistic limit of de Sitter \cite{Axenides} and anti-de Sitter spaces (usually called Newton-Hooke spaces \cite{NewtonHook}), just dS or AdS, respectively. \\
\indent The system under study in the original formulation \cite{Antonov,Bell-Wood} is a self-gravitating gas in the Newtonian limit bound by a spherical shell with insulating and perfectly reflecting walls with fixed energy (microcanonical ensemble) and fixed number of point particles (stars). An equilibrium state corresponding to an entropy maximum is called an `isothermal sphere'. Antonov proved that for such a system in the mean field approximation there is no global entropy maximum. However, there exist local entropy maxima (i.e. metastable states) for $E\cdot R > -0.335 GM^2$. This means that for positive energy $E$, there are isothermal spheres for every radius $R$, but for negative energy local entropy extrema exist only for radii smaller than a critical value $R_A$, we call Antonov radius. For $R > R_A$ and fixed negative energy there does not exist any equilibrium state. \\
\indent For a fixed negative energy and for radii smaller than the Antonov radius $R < R_A$, the equilibrium state may be stable (local entropy maxima) or unstable (saddle points) depending on the value of the ratio of core density $\rho_0 = \rho(0)$ versus the edge ratio $\rho_R=\rho(R)$. There is a critical value $(\rho_0/\rho_R)_{cr} = 709$ at which an instability sets in, i.e. for $\rho_0/\rho_R > 709$. A remarkable feature of gravity is that this instability in the microcanonical ensemble sets in when the specific heat goes from negative to positive values and not the other way around (non-equivalence of ensembles).  \\
\indent We find that gravothermal catastrophe (described in the last two paragraphs) depends crucially on the vacuum background. AdS space destabilizes the system. Compared to the flat case, the instability sets in at smaller radius, at higher central density and occurs not only for negative but also for positive energies. In dS space the phenomenon is drastically altered. A series of equilibrium solutions is restored for large radii, indicating a type of `reentrant' behaviour. A metastable homogeneous solution that suffers a transition to Antonov instability is found, together with metastable states that do not suffer any transition to instability. In addition, centrally diluted and periodically condensed solutions are allowed.
\section{Analysis}
\begin{figure}[tb!]
\begin{center}
	\subfigure[$R < R_H$]{ \label{fig:1a}\includegraphics[scale=0.5]{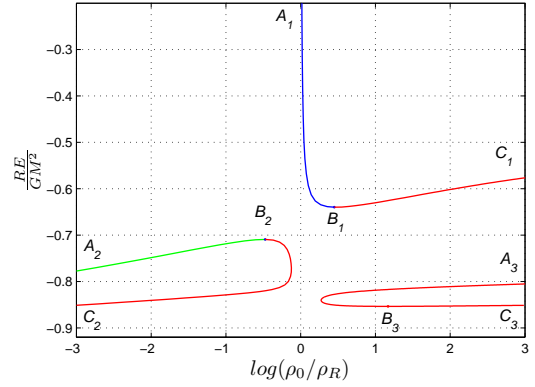} } 
\\
	\subfigure[$R > R_H$]{ \label{fig:1b}\includegraphics[scale=0.5]{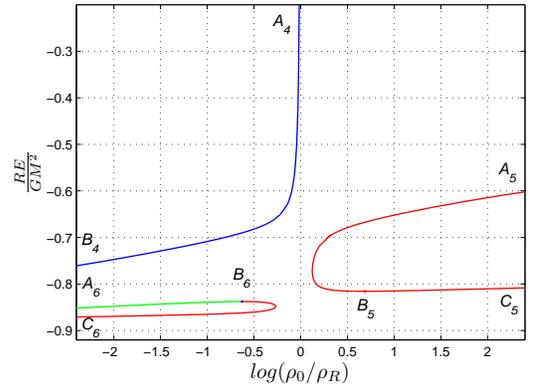} }
\caption{In dS, the dimensionless energy $Q=RE/GM^2$ versus $\log(\rho_0/\rho_R)$ for fixed $M$, $R$, $\rho_\Lambda$. Every point of any curve corresponds to a distinct isothermal sphere, so that a curve represents a series of solutions. Unlike flat case, this plot \textit{cannot} be interpreted with varying $R$ and fixed $E$, because at each point, $m$ is held constant. There exist many other curves not shown at each case for more negative energies. Only the upper series correspond to $\rho(r)$ monotonically changing. Distinct series corresponding to more negative energies have more extrema of $\rho(r)$. At points $B_1$, $B_2$, $B_3$, $B_5$ and $B_6$ an instability sets in, while curves $A_3B_3$ and $A_5B_5$ are already unstable solutions.
\label{fig:mult}}
\end{center}
\end{figure}

Consider a self-gravitating gas of $N$ point particles, each of mass $\tilde{m}=1$, inside a spherical shell with insulating and perfectly reflecting walls in the presence of a cosmological constant \cite{devega}. We work in the mean field approximation and use the one body distribution function $f(\vec{r},\vec{p})$. This is viable as long as the correlations between the particles are not significant \cite{Padman,Katz,binney}. The form of the distribution function at the equilibrium is found by the extremization of the Boltzmann entropy $S = -k\int{f \log f d^3\vec{r}d^3\vec{p}}$ under the constraints of constant energy $E$ and number of particles $N$ (i.e. constant mass $M = N\cdot \tilde{m}$). It is easy to generalize for AdS and dS vacua, Antonov's proof that a global entropy maximum does not exist. Only local entropy extrema do exist. In Ref. \cite{Antonov,Bell-Wood} is proved for the flat case that only spherical distributions maximize the entropy. In this work we consider only spherical distributions \cite{vir_appr}. We find that at an equilibrium the distribution function is: 
$f(r,\upsilon) = (\beta/2\pi)^{3/2} e^{-\beta\upsilon^2/2} \rho(r)$
, where
\begin{equation}\label{eq:f}
	\rho(r) = \rho_0 e^{-\beta(\phi - \phi(0))}
\end{equation}
is the density of matter and $\phi(r)$ is the potential that obeys the equation \cite{Wald}
\begin{equation}\label{eq:poisson}
	\nabla^2 \phi = 4\pi G \rho - 8\pi G \rho_\Lambda
\end{equation}
where the density $\rho_\Lambda$ is related to the cosmological constant by the equation $\rho_\Lambda = \Lambda c^2/8\pi G$. The potential can therefore be written as $\phi = \phi_N + \phi_\Lambda$ with:
\begin{equation}\label{eq:phi}
	\phi_N = -G\int{\frac{\rho(r')}{|\vec{r}-\vec{r}\, '|}d^3\vec{r}\, '} \; , \;
	\phi_\Lambda = -\frac{4\pi G}{3}\rho_\Lambda r^2	
\end{equation}
Using the spherical symmetry and the dimensionless variables $y = \beta(\phi - \phi (0))$, $x = r\sqrt{4\pi G \rho_0\beta}$ and $\lambda = 2\rho_\Lambda/\rho_0$ equation \eqref{eq:poisson} becomes: 
\begin{equation}\label{eq:emdenL}
	\frac{1}{x^2}\frac{d}{dx}\left( x^2\frac{d}{dx}y\right) = e^{-y} - \lambda
\end{equation}
which we call the `Emden-$\Lambda$' equation. This equation indicates that for every $\rho_\Lambda > 0$, there exists a radius $R_H$ for which a series of homogeneous solutions, with $\rho = 2\rho_\Lambda = const.$, $\phi(r)=\phi'(r)=0$, exists. This radius depends only on $\rho_\Lambda$ with $R_H = (\frac{3M}{8\pi \rho_\Lambda})^\frac{1}{3}$ (independent of energy). The homogeneous solution resembles the Einstein's static universe in the non-relativistic limit. \\
\indent In general, we want to solve the Emden-$\Lambda$ equation with initial conditions $y(0)=y'(0)=0$ for various isothermal spheres, characterized by the radius $R$, keeping $M$ constant. Let call $z = R\sqrt{4\pi G \rho_0\beta}$ the value of $x$ at $R$. The cosmological constant introduces a mass scale 
$M_\Lambda = \frac{4}{3}\pi R^3 \rho_\Lambda$ and we define the dimensionless mass $m = M/2M_\Lambda$. Using the dimensionless temperature $\bar{\beta} = GM\beta/R$, $m$ can be written as 
\begin{equation}\label{eq:m}
	m = 3\bar{\beta}/\lambda z^2
\end{equation}
Dimensionless temperature can be calculated at $z$, by integrating the Emden-$\Lambda$ equation:
\begin{equation}\label{eq:beta}
	\bar{\beta}(z) = z y'(z) + \frac{1}{3}\lambda z^2
\end{equation}
In order to keep $m$ fixed for various $z$ we see from \eqref{eq:m} that $\lambda$ has to be different at each $z$. We developed an algorithm \cite{toapp} that calculates the right $\lambda$ for each $z$ and solves Emden-$\Lambda$ with respect to $z$, keeping $m$ constant. Writing $m$ as $m = \frac{3}{8\pi}\frac{M}{\rho_\Lambda R^3}$ it is clear that solving for various fixed $m$ can be interpreted as solving for various $\rho_\Lambda$ and/or $R$ for a fixed $M$. \\
\indent Using the virial \cite{toapp,vir_appr} theorem:  $2K+U_N - 2U_\Lambda = 3PV
$ we derive the following expression for the non-dimensional energy at $z$, $Q(z) \equiv RE/GM^2$:
\begin{equation}\label{eq:en}
	Q(z) = \frac{z^2 e^{-y(z)}}{\bar{\beta}^2} - \frac{3}{2\bar{\beta}} - \frac{\lambda}{2\bar{\beta}^2 z}\int_0^z{x^4 e^{-y(x)}dx}
\end{equation}
Using \eqref{eq:en} $Q$ can be calculated numerically. For dS it is drawn in Figure \ref{fig:mult}. \\
\indent In order to conclude on the type of stability we use Poincar\'e's linear series of equilibria theorem \cite{Bell-Wood,Poincare,Katz}. It states that as $S(E)$ varies from one equilibrium to another by infinitesimal changing the parameter $E$ (with $E$ being held fixed in order to calculate the extremum of $S$), the type of the extremum does change if this linear series of equilibria meets another series or turns back through previous values of $E$. The later is our case. Therefore at an extremum of $E$, the stability changes. Complementary to this criterion we use the second variation of entropy (following closely Padmanabhan \cite{Padman}):
\begin{equation}\label{eq:dSK}
	\delta^2S = \int_0^R\int_0^R \delta M(r_2) \hat{K}(r_1,r_2) \delta M(r_1) dr_1 dr_2
\end{equation}
where $\delta M$ is a local mass perturbation and:
\begin{eqnarray}\label{eq:K}
	\hat{K}(r_1,r_2) = &-&\frac{\phi'(r_1)\phi'(r_2)}{3MT^2} +\frac{1}{2}\delta(r_1-r_2)\nonumber\\
	&\times& \left[ \frac{G}{Tr_1^2} + \frac{d}{dr_1}\left(\frac{1}{4\pi\rho r_1^2} \frac{d}{dr_1}\right)\right]
\end{eqnarray}
The sign of $\delta^2 S$ is determined by the sign of the eigenvalues $\xi$ of the eigenvalue problem:
\begin{equation}\label{eq:eig}
	\int_0^R \hat{K}(r,r_1) F_\xi(r_1) dr_1 = \xi F_\xi(r) 
\end{equation}
with $F_\xi(0) = F_\xi(R) = 0$.
We developed an algorithm that can numerically determine eigenvalues and eigenstates (the perturbations) of equation \eqref{eq:eig}. If, for an equilibrium, equation \eqref{eq:eig} has solutions with one or more positive eigenvalues $\xi$, then this equilibrium is unstable, because for this mode $\delta^2 S > 0$. \\
\indent The instability of an equilibrium can be proved by finding just one unstable mode. However, the stability is much harder to be proved. A strategy is to prove stability for a limiting case and claim that the stability does not change across various $E$ until $E$ reaches an extremum (Poincar\'e). We verified numerically this claim for the specific solutions presented here. In the flat case, stability is proved \cite{Antonov, Padman} for $R\rightarrow 0$. In this limit the presence of a cosmological constant is irrelevant. Thus, the same argument holds for AdS and for dS only when $R<R_H$ for the curve $A_1B_1$ of Figure \ref{fig:1a}. If $R > R_H$ this argument does not hold and we prove stability for the curve $A_4B_4$ of Figure \ref{fig:1b} as follows: the limit $\beta \rightarrow 0$ does exist for these series of equilibria\footnote{
Intuitively, one can argue that for $\beta\rightarrow 0$ the kinetic energy dominates, therefore the system behaves as an ideal gas, hence it is thermodynamically stable.} since it corresponds to $E\rightarrow\infty$. We see that $\delta^2 S < 0$ for $\beta \rightarrow 0$, when $\delta^2 S$ is written in the form:
\begin{eqnarray}\label{eq:2entr}
	\delta^2S = &-&\frac{\beta^2}{3M}\left(\int_0^R\phi'\delta M dr \right)^2  +\frac{G\beta}{2}\int_0^R \frac{(\delta M)^2}{r^2}dr \nonumber \\
	&-& \int_0^R \frac{((\delta M)')^2}{8\pi\rho r^2}dr
\end{eqnarray}
Note that in equations \eqref{eq:dSK}, \eqref{eq:2entr} the cosmological constant enters implicitly through the derivative of the potential. \\
\indent For the homogeneous solution at $R = R_H$ we find that although $E$ has not an extremum, there exists a critical point $\delta^2S = 0$ determined analytically, with $z_{cr} = 4.493 \Leftrightarrow \bar{\beta}_{cr} = 6.73$, where the stable series of equilibria become unstable. It is evident that such a point should exist, since $\beta \rightarrow 0$ leads to a stable state (equation \eqref{eq:2entr}), while $\beta \rightarrow \infty$ gives the unstable static state. 
\begin{figure}[tb!]
\begin{center}
	\subfigure[]{ \label{fig:2a}\includegraphics[scale=0.39]{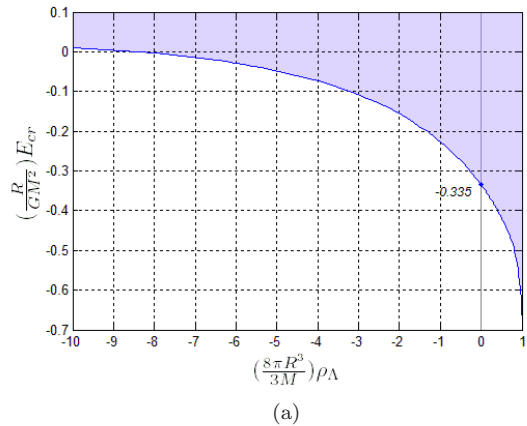} } 
\\
	\subfigure[]{ \label{fig:2b}\includegraphics[scale=0.5]{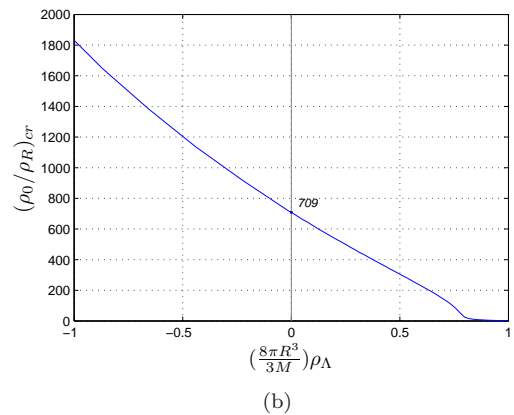} }
\caption{The critical energy and critical ratio of the centre to edge density for fixed radius $R$ and fixed mass $M$ vs the cosmological constant. (a)For $E < E_{cr}$ (unshaded region) there is no equilibrium state, while in the shaded region there exist equilibria (stable or unstable, depending on the ratio $\rho_0/\rho_R$). (b)For $\rho_0/\rho_R < (\rho_0/\rho_R)_{cr}$ the equilibrium is stable, while for $\rho_0/\rho_R > (\rho_0/\rho_R)_{cr}$ it is unstable.
\label{fig:ER}}
\end{center}
\end{figure}
\begin{figure}[tb]
\begin{center}
	\includegraphics[scale=0.40]{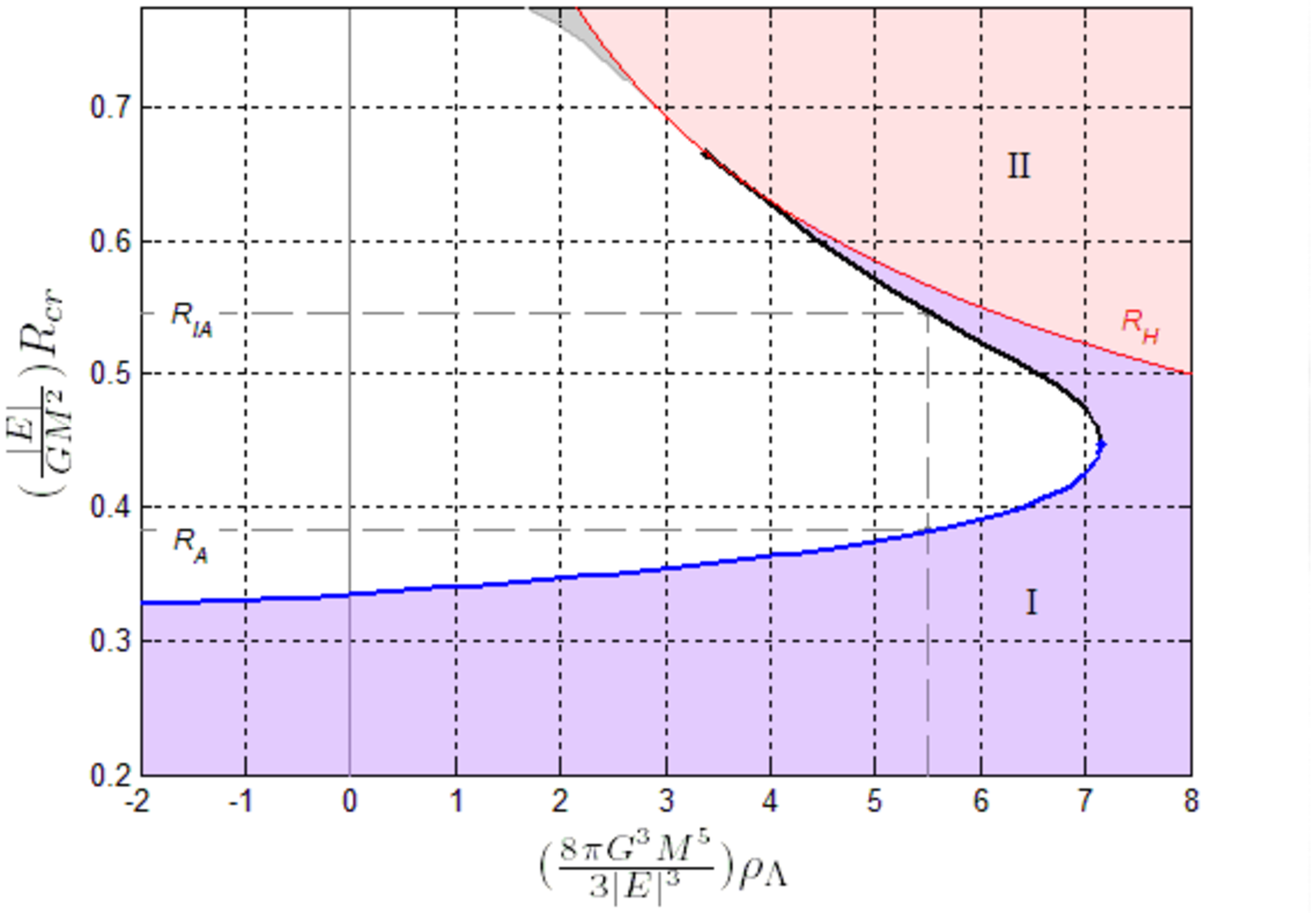} 
\end{center}
\caption{The critical radius $R_{cr}$ for a fixed negative energy $E$ and mass $M$ versus the cosmological constant $\rho_\Lambda$. For a given large enough and positive $\rho_\Lambda$ there exist two critical radii $R_A$ and $R_{IA}$. In the white unshaded area there exist no equilibrium states. The line $R=R_H$ is the radius for each $\rho_\Lambda$ where a homogeneous $\rho = 2\rho_\Lambda = const.$ solution exists. In the region I there exist equilibria for which $\rho(r)$ is monotonically decreasing that are metastable or unstable states, depending on the ratio $\rho_0/\rho_R$. In the region II there exist metastable states for which $\rho(r)$ is monotonically increasing that do not suffer a transition to Antonov instability. The small gray shaded region contains equilibria with $\rho(r)$ not monotonic with one or more maxima and $\rho(r) > \rho_0$. This region stops at $\rho_\Lambda^{min} = \bar{\rho}/4$.} 
\label{fig:Rcr}
\end{figure}
\begin{figure}[tb]
\begin{center}
	\includegraphics[scale=0.533]{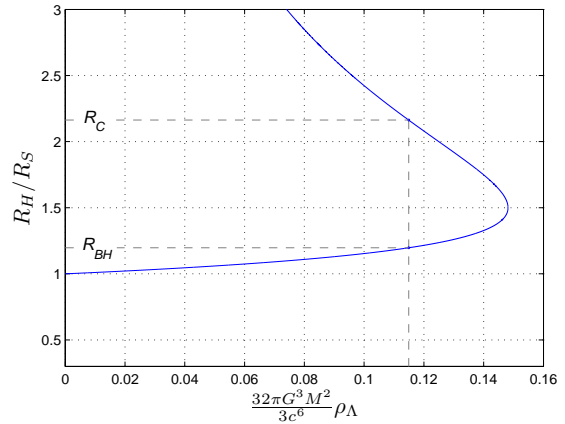} 
\end{center}
\caption{Schwartzschild-dS space has two horizons, a black hole horizon $R_{BH}$ and a cosmological horizon $R_C$. In this figure,  the horizon radius $R_H$ is measured in units of the Schwartzschild radius $R_S = \frac{2GM}{c^2}$ and is plotted w.r.t. the cosmological constant for a fixed mass.} 
\label{fig:SdS}
\end{figure}
\section{Results}
For a fixed radius $R < R_A$, where metastable states with negative energy do exist, we find that the critical energy $E_{cr}$ and the critical ratio $(\rho_0/\rho_R)_{cr}$ are decreasing with increasing cosmological constant $\rho_\Lambda$ as one can see in Figure \ref{fig:ER}. $E_{cr}$ is the minimum energy for which an isothermal sphere exists. For AdS the instability at $(\rho_0/\rho_R)_{cr}$ sets in not only for negative, but also for positive energies, because the potential energy of matter coupled to AdS is positive. In Figure \ref{fig:Rcr} can be seen the critical radius $R_{cr}$ versus the cosmological constant $\rho_\Lambda$. As AdS becomes stronger, the region of no equilibrium ($R > R_A$) is getting bigger. All these facts  further support the recent investigations of the newly discovered non-linear instability of AdS spacetime \cite{AdS1,AdS2}. \\
\indent  The de Sitter case is more complicated, though very interesting features arise. The critical Antonov radius $R_A$ increases with increasing $\rho_\Lambda$ defining the blue `Antonov line' in Figure \ref{fig:Rcr}. 
There also appears a second line, where an `inverse Antonov transition' occurs (black line in Figure \ref{fig:Rcr}) 
defined by a second critical radius $R_{IA}$. This radius is decreasing with increasing $\rho_\Lambda$. The two critical radii merge 
at an extremal value $\rho^{max}_\Lambda$. This is a typical `\textit{reentrant}' behaviour. Reentrant phenomena of this type are common in statistical systems, whenever competing interactions are present \cite{reentrant1,reentrant2,reentrant3,reentrant4}. For a fixed $\rho_\Lambda$, for $R < R_A$ metastable states do exist and for $R_A < R < R_{IA}$ no thermodynamic equilibria exist. However, for $R>R_{IA}$ the system `reenters' to a phase, where metastable states are possible. dS acts as a stabilizer for the system, since as dS becomes stronger the region of no equilibrium diminishes. For $\rho_\Lambda > \rho^{max}_\Lambda$ there are metastable states for every $R$ and fixed $E$. \\
\indent One may intuitively understand the existence of a second critical radius $R_{IA}$ as follows. It is this radius beyond which the cosmological force that is exerted on the outer parts is so powerful that can hold enough mass at the outer regions, so that the remaining matter in the inner regions shall not collapse. \\ 
\indent As already mentioned in the analysis, the homogeneous solution is stable for temperature $T > T_{cr}$ , while for $T < T_{cr}$ is unstable, with $T_{cr} = \frac{GM}{6.73R_H}$. Apart from the homogeneous solution, for $R = R_H$ there exist infinitely many other series of equilibria with various configurations $\rho(r)$. These with $\rho(r) < \rho_0$ are unstable, while these with $\rho(r) > \rho_0$ are stable. \\
\indent For $R < R_H$, in region I of Figure \ref{fig:Rcr}, there exist metastable states which correspond to the line $A_1B_1$ of Figure \ref{fig:1a}. For these, $\rho(r)$ is monotonically decreasing and can become unstable depending on the value of $\rho_0/\rho_R$ likewise flat case. In the small upper gray shaded region of Figure \ref{fig:Rcr} there exist equilibria with $\rho(r)$ \textit{not} monotonic, with one maximum and $\rho(r) > \rho_0$, $\forall r$. These correspond to the curve $A_2B_2$ of Figure \ref{fig:1a}. Numerical calculations indicate that this curve corresponds to stable solutions. At the maximum of energy $B_2$ an instability sets in. There exist a lot of other series with more maxima of $\rho(r)$ that are not drawn in Figure \ref{fig:mult}. All these solutions correspond to isothermal spheres diluted at the center and with concentric condensations away from the center. \\
\indent At the region $R > R_H$ (region II in Figure \ref{fig:Rcr}) there exist metastable states with $\rho(r)$ monotonically increasing (curve $A_4B_4$ in Figure \ref{fig:1b}), that do not suffer a transition to instability. These isothermal spheres are somewhat hollow with the mass being mostly concentrated at the edge. In addition there exist all other solutions shown in Figure \ref{fig:1b}. \\
\indent The small upper gray shaded region of Figure \ref{fig:Rcr} contains the solutions $A_2B_2$ of Figure \ref{fig:1a} along with other similar solutions for more negative energy that cannot be seen in Figure \ref{fig:1a}. All these, are finite in number for a fixed $\rho_\Lambda$. Their multitude is diminishing as $\rho_\Lambda$ is getting smaller. They exist only for $\rho_\Lambda > \rho_\Lambda^{min}$ with $\rho_\Lambda^{min} = \bar{\rho}/4$, where $\bar{\rho}$ is the mean density. Therefore, the whole small gray region of Figure \ref{fig:Rcr}, stops at $\rho_\Lambda^{min}$ and its lower boundary does not cross the orthogonal axis. \\
\indent The value $\rho_\Lambda^{min}$ can be calculated as follows. The density contrast of series $A_2B_2$ behaves as $\rho_0/\rho_R\rightarrow 0$ for diminishing cosmological constant. The $\rho_\Lambda^{min}$ is this value, for a fixed radius and mass, for which the cosmological force is so strong that marginally can hold the entire mass at the edge, i.e. $GM/2R^2 \leq \frac{8\pi G}{3}\rho_\Lambda R$ $\Rightarrow$ $\rho_\Lambda \geq 3M/16\pi R^3$ so that $\rho_\Lambda^{min} = 3M/16\pi R^3 = \bar{\rho}/4$. This result is supported by the numerical calculations, as well. \\
\indent Let us discuss the validity of the Newtonian limit with a cosmological constant (see Appendix, as well). In order to meet for $\Lambda$ the weak field condition, required by the Newtonian approximation, the following condition is imposed when taking the Newtonian limit in Einstein's equations with $\Lambda$.
\begin{equation}\label{eq:LApp}
	\frac{1}{3}\Lambda R^2 \ll 1
\end{equation}
It is straightforward to verify that this condition leads to the condition
\begin{equation}\label{eq:mapp}
	m \gg \frac{1}{2}\frac{R_S}{R}
\end{equation}	
where $m = 3M/8\pi\rho_\Lambda R^3$ is defined in equation (\ref{eq:m}) and $R_S = 2GM/c^2$ is the Schwartzshild radius of the system. Condition (\ref{eq:mapp}) defines the validity of Newtonian approximation. The value of $m$ that corresponds to the extremal $\rho_\Lambda^{max}$ of Figure \ref{fig:Rcr} is $m \simeq 1.57$, satisfying perfectly condition (\ref{eq:mapp}), along with all other points of $R_{cr}$ of Figure \ref{fig:Rcr}, since they have $m>0.9$. Another issue that might be raised is whether in general, $\Lambda$ is negligible for Newtonian systems. That is not true. From Poisson with $\Lambda$ equation (\ref{eq:poisson}) we see that the cosmological constant is negligible for a Newtonian system, if
\begin{equation}\label{eq:Lneg}
	\rho \gg \rho_\Lambda \Leftrightarrow \frac{1}{3}\Lambda R^2 \cdot \frac{R}{R_S} \ll 1 
\end{equation} 
Apparently, equation (\ref{eq:LApp}) does not necessarily imply equation (\ref{eq:Lneg}). It may very well happen that condition (\ref{eq:LApp}) holds with $\Lambda R^2 \cdot (R/R_S) \sim  1 $ for systems with $R/R_S \gg 1$. \\
\indent An example from the physical world, when condition (\ref{eq:LApp}) holds but not condition (\ref{eq:Lneg}) is galaxy clusters and superclusters. There is evidence that the cosmological constant plays a role to the evolution of the galaxy clusters and superclusters \cite{Axenides,Voit}. In addition gravothermal catastrophe on its own, could play a role, as well, since many present a core-halo structure and others have a supermassive black hole in their centres. For some typical values \cite{Voit,Mathews} of $M\simeq 10^{15}M_\odot,R\simeq 2-5Mpc,E\simeq -10^{57}J$ of regular galaxy clusters, their mean density $\bar{\rho}$ is of the same order of magnitude as the cosmological constant $\simeq 10^{-26}kgr/m^3$. Thus $2\rho_\Lambda/\bar{\rho}$, that is the variable of the horizontal axis of Figure \ref{fig:ER}, is of order unity. For the same typical values, the quantity $8\pi G^3 M^5\rho_\Lambda/3|E|^3$, i.e. the variable of the horizontal axis of Figure \ref{fig:Rcr}, is of order unity, as well. This is evidence that our analysis could be relevant to the evolution of galaxy clusters. We believe that the effect of gravothermal catastrophe to the evolution of galaxy clusters on its own, and secondly the possible impact of the cosmological constant on it, deserve further investigation. \\
\indent Last but not least, let us stress out that the reentrant behaviour of Figure \ref{fig:Rcr} resembles the two horizons of the Schwartzschild-dS black hole in General Relativity. The radii of these horizons are plotted in Figure \ref{fig:SdS} with respect to the cosmological constant. The similarity between Figure \ref{fig:Rcr} and Figure \ref{fig:SdS} is too striking to be considered a coincidence! It seems that the reentrant phenomenon we have discovered is the closest Newtonian analogue to the two horizons of Schwartzschild-dS space. However, there is a big difference. The stable region of the Newtonian case corresponds to the unstable region of the relativistic case. This opposite sense, with no cosmological constant present, is explained in Ref. \cite{Chavanis3}. 

\begin{acknowledgements} 
We thank C. Efthymiopoulos and E.G. Floratos for useful comments and especially S. Nicolis for discussions on the reentrance phenomenon.
\end{acknowledgements}

\appendix*

\section{}
We analytically derive from General Relativity the Emden with $\Lambda$ equation and discuss the validity of the Newtonian approximation. \\
\indent The Emden equation is well known to be the Newtonian limit of the Tolman-Oppenheimer-Volkov (TOV) equation for dust (non-relativistic) particles. We will show that similarly the Emden-$\Lambda$ equation is the Newtonian limit of the TOV equation with $\Lambda$. \\
\indent The relativistic TOV equation in the presence of $\Lambda$ is:
\begin{eqnarray}\label{eq:TOVL}
	\frac{dp}{dr} = &-&(p/c^2 + \rho)\left(\frac{GM(r)}{r^2} + \frac{4\pi G}{c^2}p\cdot r - \frac{1}{3}\Lambda c^2 r \right) \nonumber\\ 
	&\times&\left(1 - \frac{2GM(r)}{r c^2} - \frac{1}{3}\Lambda r^2 \right)^{-1}
\end{eqnarray}
with 
\begin{equation}\label{eq:MASSC}
	\frac{dM(r)}{dr} = 4\pi\rho r^2
\end{equation}
For a perfect fluid the equation of state is $p = w\rho c^2$ with $w = \frac{kT}{\tilde{m}c^2}$, where $\tilde{m}$ is the rest mass of one particle. For non-relativistic particles it is $w\rightarrow 0$, i.e.
\[
	kT \ll \tilde{m}c^2
\]
Substituting equation $p = w\rho c^2$ into the TOV-$\Lambda$ equation (\ref{eq:TOVL}) and taking the dust (non-relativistic) particles limit $kT \ll \tilde{m}c^2$, we get
\begin{equation}\label{eq:TOVL_NR}
	\frac{kT}{\tilde{m}}\frac{d\rho}{dr} = -\rho\left(\frac{GM(r)}{r^2} - \frac{1}{3}\Lambda c^2 r \right)\cdot\left(1 - \frac{1}{3}\Lambda r^2 \right)^{-1}
\end{equation}
Note that the term $(2GM(r)/r)kT/\tilde{m}c^2$ is negligible so that 
in general relativity with $\Lambda = 0$ just the limit $w\rightarrow 0$ gives the Emden equation \cite{Chandra}, provided there is no singularity. \\
\indent In order to obtain the Newtonian limit one has to assume that the gravitational field related to the cosmological constant is weak. We see from the above equation
(\ref{eq:TOVL_NR}) that this accounts for:
\begin{equation}\label{eq:NLIMIT}
	\frac{1}{3}\Lambda R^2 \ll 1 \quad \text{, Newtonian limit}
\end{equation}
since $r \leq R$. In this limit and substituting $\rho_\Lambda = \Lambda c^2/8\pi G$, $\beta = 1/kT$ equation (\ref{eq:TOVL_NR}) becomes
\begin{equation}\label{eq:TOVL_NR_1}
	r^2\frac{d(-\frac{1}{\beta \tilde{m}}\log\rho)}{dr} = GM(r) - \frac{1}{3}8\pi G\rho_\Lambda r^3
\end{equation}
which after differentiation w.r.t $r$ and substituting equation (\ref{eq:MASSC}) gives finally:
\begin{equation}\label{eq:NEWT}
\frac{1}{r^2}\frac{d}{dr}\left(r^2\frac{d\phi}{dr} \right) = 4\pi G \rho - 8\pi G \rho_\Lambda
\end{equation}
with $\rho = A e^{-\beta \tilde{m}\phi}$. This is the Emden-$\Lambda$ equation. It is evident that the cosmological constant is negligible if
\begin{equation}\label{eq:LNEGL}
\rho \gg \rho_\Lambda \Leftrightarrow \frac{1}{3}\Lambda R^2\cdot \frac{R}{R_S} \ll 1 \quad \text{, negligible $\Lambda$}
\end{equation}
where $R_S = 2GM/c^2$ is the Schwartzscild radius. The equation (\ref{eq:NLIMIT}) for the Newtonian limit does not necessarily imply equation (\ref{eq:LNEGL}) for negligible $\Lambda$ since for Newtonian systems it normally holds that $R/R_S \gg 1$.



\begin{thebibliography}{999}

\bibitem{Antonov}
  V.A. Antonov,
  Vest. Leningrad Univ. {\bf 7}, 135, (1962)

\bibitem{Bell-Wood}
  D. Lynden-Bell and R. Wood,
  MNRAS {\bf 138}, 495, (1968)

\bibitem{Padman}
  T. Padmanabhan, 
  Phys. Rep. {\bf 188}, 285, (1990)

\bibitem{Chavanis}
  P.H. Chavanis, 
	A{\&}A {\bf 381}, 340, (2002)

\bibitem{Bell}
	D. Lynden-Bell,
	Physica A {\bf 263}, 293, (1999)

\bibitem{Dauxois}
	A. Campa, T. Dauxois and S. Ruffo, 
	Phys. Rep. {\bf 480}, 57 (2009)

\bibitem{Chavanis2}
  P.H. Chavanis, 
  Int. J. Mod. Phys. B, {\bf 20}, 3113 (2006)

\bibitem{Padman2}
  T. Padmanabhan, 
  Phys. Rep. {\bf 380}, 235, (2003)
  
\bibitem{Harvey}
  A. Harvey,
  Eur. J. Phys. {\bf 30}, 877, (2009)

\bibitem{AdSCFT}
  O. Aharony, S.S. Gubser, J. Maldacena, H. Ooguri, and Y. Oz, 
  Phys. Rep. {\bf 323}, 183, (2000)

\bibitem{AdS1}
  P. Bizo{\'n} and A. Rostworowski, 
  Phys. Rev. Lett. {\bf 107}, 031102, (2011)

\bibitem{AdS2}
  O. J.C. Dias, G.T. Horowitz, and J.E. Santos,
  Classical Quantum Gravity {\bf 29}, 194002 (2012)

\bibitem{Axenides}
	M. Axenides, E.G. Floratos and L. Perivolaropoulos,
	Mod. Phys. Lett. {\bf A15}, 1541, (2000) 

\bibitem{NewtonHook}
	G.W. Gibbons and C.E. Patricot,
	Classical Quantum Gravity {\bf 20}, 5225, (2003)

\bibitem{devega}
  H.J. de Vega and J.A. Siebert,
  Nucl. Phys. B {\bf 707}, 529, (2005)

\bibitem{Katz}
  J. Katz, 
  Found. Phys. {\bf 33}, 223, (2003)

\bibitem{binney}
  J. Binney and S. Tremaine,
  `Galactic Dynamics', Princeton (1987)
  
\bibitem{vir_appr}
  M. Nowakowski, J.C. Sanabria, A. Garcia,
  Phys. Rev. D {\bf 66}, 023003, (2002)

\bibitem{Wald}
	S.R. Green and R.M. Wald,
	Phys. Rev. D {\bf 85}, 063512, (2012)

\bibitem{toapp}
	M. Axenides, G. Georgiou and Z. Roupas,
	to appear

\bibitem{Poincare}
  H. Poincar{\'e}, 
  Acta. Math. {\bf 7}, 259, (1885)

\bibitem{reentrant1}
	F. Staniscia, P.H. Chavanis, De Ninno and D. Fanelli,
	Phys. Rev. E {\bf 80}, 021138, (2009) 

\bibitem{reentrant2}
	C.K. Thomas and H.G. Katzgraber,
	Phys. Rev. E {\bf 84}, 040101(R), (2011)

\bibitem{reentrant3}
	T. Dauxois, P. de Buyl, L. Lori and S. Ruffo,
	J. Stat. Mech. P06015, (2010) 

\bibitem{reentrant4}
	F. Staniscia, P.H. Chavanis and G. De Ninno,
	Phys. Rev. E {\bf 83}, 051111 (2011) 

\bibitem{Voit}		
	G. Mark Voit,
	Rev. of Mod. Phys., {\bf 77}, 207 (2005)	

\bibitem{Mathews}		
	William G. Mathews and Fulai Guo, 
	ApJ, {\bf 738}, 155 (2011)	
	
\bibitem{Chavanis3}
  P.H. Chavanis, 
	A{\&}A, {\bf 381}, 709 (2002)
	
\bibitem{Chandra}
	S. Chandrasekhar,
	`A limiting case of relativistic equilibrium', 
	\textit{ General Relativity} p. 185-199,
	Edited by L.O' Raifeartaigh, Oxford, (1972)
	
\end{thebibliography}
\end{document}